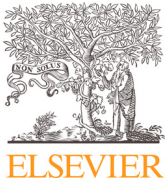
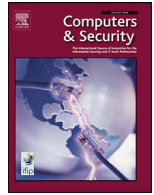

# A hands-on gaze on HTTP/3 security through the lens of HTTP/2 and a public dataset

Efstratios Chatzoglou[a], Vasileios Kouliaridis[a], Georgios Kambourakis[a], Georgios Karopoulos[b,∗], Stefanos Gritzalis[c]

[a] *Department of Information and Communication Systems Engineering, University of the Aegean, Karlovasi 83200, Samos, Greece*
[b] *European Commission, Joint Research Centre, Ispra 21027, Italy*
[c] *Department of Digital Systems, University of Piraeus, Piraeus 18532, Greece*

**ABSTRACT**

Following QUIC protocol ratification on May 2021, the third major version of the Hypertext Transfer Protocol, namely HTTP/3, was published around one year later in RFC 9114. In light of these consequential advancements, the current work aspires to provide a full-blown coverage of the following issues, which to our knowledge have received feeble or no attention in the literature so far. First, we provide a complete review of attacks against HTTP/2, and elaborate on if and in which way they can be migrated to HTTP/3. Second, through the creation of a testbed comprising the at present six most popular HTTP/3-enabled servers, we examine the effectiveness of a quartet of attacks, either stemming directly from the HTTP/2 relevant literature or being entirely new. This scrutiny led to the assignment of at least one CVE ID with a critical base score by MITRE. No less important, by capitalizing on a realistic, abundant in devices testbed, we compiled a voluminous, labeled corpus containing traces of ten diverse attacks against HTTP and QUIC services. An initial evaluation of the dataset mainly by means of machine learning techniques is included as well. Given that the 30 GB dataset is made available in both pcap and CSV formats, forthcoming research can easily take advantage of any subset of features, contingent upon the specific network topology and configuration.

## 1. Introduction

HTTP was originally designed without focusing on security and reliability; this is one of the main motivations behind the development of HTTP/2 Thomson and Benfield (2022). However, as we discuss in detail in Section 2, the adoption of HTTP/2 introduced new attacks, as happened also in the past with the rather quick release of novel technologies that were later found to have security issues Chatzoglou et al. (2021, 2022b); Karopoulos et al. (2021b); Kouliaridis et al. (2021).

The next major HTTP version, namely HTTP/3 Bishop (2022), is an upgrade of HTTP/2 in terms of performance, reliability, and security; at the same time, it is based on the QUIC protocol Iyengar and Thomson (2021) and it heavily changes the way web browsers and servers communicate, given that it uses UDP as a transport layer protocol instead of TCP, making it a candidate source of further security issues.

Considering also stagnating security issues of HTTP, such as the low penetration rate of HTTP security headers Karopoulos et al. (2021a) (below 17% across all platforms), as well as VNC mr.d0x (b); Tommasi et al. (2022) and iframe mr.d0x (a); Tommasi et al. (2022) phishing attacks, the following question arises: what is the security status of the new generations of HTTP, that is, HTTP/2 and HTTP/3?

Deployment-wise, according to W3Techs (2022), HTTP/2 has currently an adoption rate of 45.2%, which is at about the same level as one year ago (45.4%). In between, this rate went up to a maximum of 46.9% in Jan. 2022, following a declining trajectory ever since. HTTP/3, on the other hand, followed a steady upward adoption rate from 19.5% one year ago to 25% in Jun. 2022. It is also noteworthy that a new candidate was added to the existing ones for supporting encrypted DNS Kambourakis and Karopoulos (2022), that is, DNS over HTTP/3 or DoH3 Google (2022). These data indicate that the adoption of HTTP/2 is relatively stable, but losing ground and HTTP/3 is taking its place, albeit in a slow pace.

∗ Corresponding author.
*E-mail addresses:* efchatzoglou@aegean.gr (E. Chatzoglou), bkouliaridis@aegean.gr (V. Kouliaridis), gkamb@aegean.gr (G. Kambourakis), georgios.karopoulos@ec.europa.eu (G. Karopoulos), sgritz@unipi.gr (S. Gritzalis).

This underlines the need to evaluate the security of HTTP/2, with a view to protect today's vulnerable deployments, but at the same time consider the issues that HTTP/3 will bring in the near future when it will become the dominant protocol version.

Even though a significant mass of work has been accomplished on the analysis of HTTP/2 vulnerabilities, to the best of our knowledge, no extensive review exists to provide a spherical analysis on the security of HTTP/2. In fact, existing works in this field investigate individual attacks on HTTP/2, whereas very few of them evaluate HTTP/2 against a wide variety of attacks. Moreover, no insight is provided into the applicability of HTTP/2 attacks to the latest HTTP/3 version.

The work at hand aims to address the aforementioned issues and provides the following contributions:

- A comprehensive review of HTTP/2 security and known attacks in the literature.
- A discussion on which HTTP/2 security attacks could be applicable to HTTP/3 as well.
- A hands-on evaluation of QUIC and/or HTTP/3 enabled servers against HTTP/2 and HTTP/3 attacks.
- A state-of-the-art dataset built to evaluate HTTP/2, HTTP/3, and QUIC security, as well as a thorough evaluation of the proposed dataset, mainly by means of machine learning techniques.

The paper is organized as follows. The next section surveys several types of attacks on HTTP/2 and discusses their portability to HTTP/3. Section 3 provides an evaluation of QUIC and/or HTTP/3 enabled servers against common attacks. In Section 4, we present our new dataset, created specifically to assess the security of the latest HTTP protocols. Section 5 is devoted to the evaluation of the proposed dataset. The last section concludes.

## 2. Categories of attacks against HTTP/2

This section surveys the major categories of attacks against HTTP/2; moreover, the discussion focuses on if and to what degree a specific category migrates to HTTP/3. It should be noted here that previous work on web attacks Chatzoglou et al. (2022a); Gil (2017); Mirheidari et al. (2020); Noam Moshe, Sharon Brizinov, and Team82 (2022); Snyk Security Research team (2022); Tsai (2017); Web cache deception escalates (2022) has shown that server implementations are exposed to issues such as URL parsing, which may lead to server-side request forgery (SSRF) or path traversal attacks, and cache poisoning, which can enable an opponent to steal information or mount a remote code execution (RCE) attack. Additionally, works such as Patni et al. (2017); Zhang et al. (2020), illustrated different empirical attacks based on TLS vulnerabilities that could lead to MitM attacks. While the aforementioned assaults concern server-side attacks over the HTTP, they are irrelevant of the HTTP protocol version used and they are considered to be out-of-scope of this paper; thus, such attacks are omitted from the analysis that follows.

### 2.1. Amplification attacks

The work in Beckett and Sezer (2017b) examined the possibility of amplification attacks, termed HTTP/2 Tsunami, by capitalizing on the HTTP/2's HPACK header compression method. To store the requested headers in a first-in first-out fashion Internet Engineering Task Force (IETF) (b), HPACK uses a dynamic table. The authors assumed that, by exploiting HPACK, HTTP/2-enabled proxies could be used as amplifiers. To this end, they calculated the exact length of each packet header based on the length of the dynamic table, which can be assigned by the SETTINGS_HEADER_TABLE_SIZE field, having a default value 4 KB. They simultaneously sent multiple packets to the Nginx and nghttp2 proxies, with the assist of three different headers, namely, *Authority, User agent*, and *Cookie*. They resulted into having four cases with a bandwidth amplification factor of 79.2, 94.4, 140.6, and 196.3, for 100, 128, 256, and 512 maximum concurrent requests, respectively. The latter field is directly related to the dynamic table and refers to the number of simultaneous connections the server can handle at one time. The authors mentioned that they altered this field for each assault, given that this field is directly related with the amplification factor. It should be noted that the 100 max concurrent requests was the default value on each proxy.

Regarding HTTP/3, in the recently published RFC Bishop (2022), HPACK was replaced by QPACK, due to the incapability of QUIC to handle an order (first-in first-out). QPACK handles requests differently, thus, the HTTP/2 Tsunami attack is not directly applicable against HTTP/3 proxies. Nevertheless, it is interesting to examine whether the main ideas behind this attack could affect HTTP/3.

### 2.2. Cryptojacking attacks

The authors in Suresh et al. (2020) explored the feasibility of taking advantage of HTTP/2 proxies to perform cryptojacking, that is, consuming resources for mining cryptocurrencies without the consent of the resources' owner. Precisely, the attacker uses a malicious HTTP/2 proxy to downgrade the connection to a cleartext HTTP/1.1 one and inject a cryptojacking payload that the victims machine executes, starting unwillingly a cryptomining procedure.

Regarding mitigation methods, the authors suggested that such attacks can be blocked by any adblock software; on the other hand, such blocking could potentially be evaded by encrypting the cryptojacking payload with the assist of a custom stratum pool braiins.com (2022). However, it is interesting to note here that the execution of this attack, as presented in Suresh et al. (2020), is questionable; RFC 7230 Internet Engineering Task Force (IETF) (a) states that "*A server must not switch to a protocol that was not indicated by the client in the corresponding request's Upgrade header field*". In other words, a server would never initiate a protocol upgrade, but it would do so only after a client sent an upgrade request. Concerning the portability of cryptojacking to HTTP/3, we argue that this attack is based on modifying the connection to a cleartext one; given that HTTP/3 does not have a cleartext mode, the attack cannot be applied as is.

### 2.3. Denial of Service attacks

The work in Adi et al. (2016) presented a distributed denial-of-service (DDoS) attack model where malicious traffic mimics flash crowds, based on the assumption that legitimate HTTP/2 flash crowd traffic has the same network characteristics as a DDoS. Specifically, they investigated four different cases: (a) a flood-based DoS, (b) modifying the WINDOW_UPDATE size, (c) modifying the number of packets, and (d) finding the minimum number of attacking bots to mount a successful DDoS attack using the parameters found in the previous two cases. The results showed that HTTP/2 does not limit the exchanged traffic, and additional mechanisms should be devised to monitor and react to network patterns that could lead to DoS. This work is based on a similar testbed setup as Adi et al. (2015), whereas both are based on a known vulnerability on flow control and more specifically on the WINDOW_UPDATE size, which has been identified as a potential waste of resources if abused Thomson and Benfield (2022). Given that Adi et al. (2015) examined slow rate DoS attacks, it is further analyzed in Section 2.4. Notably, attacks that are related to WINDOW_UPDATE are infeasible on HTTP/3 because that field was removed from the specification Bishop (2022).

The work in Beckett and Sezer (2017a) presented an experimental analysis of the vulnerability of HTTP/1 and HTTP/2 against flood DDoS attacks. The scenario involved generating the maximum number of requests possible, first against an HTTP/1 and then against an HTTP/2 server. The results showed that, in both cases, the bottleneck of the attack was the packet generation on the attacker side due to limited processing power and offload capability of the network card. The main difference between the two experiments is that, due to multiplexing, 57 to 95 times more packets were created and sent in HTTP/2 compared to those sent in HTTP/1. This suggests that, even though HTTP/2 provides some performance benefits, it makes flood attacks more effective at the same time. Overall, this attack is a typical HTTP flooding attack, exploiting HTTP/2 characteristics to become more effective; under this prism, it is possible that a similar attack can affect HTTP/3 as well.

The work in Hu et al. (2018) examined six different attacks that could theoretically affect a 5G core network using HTTP/2 as an application layer protocol for service-based interfaces. This work is purely theoretical and does not provide any implementation of the suggested attacks. Moreover, even though not explicitly mentioned, these assaults can be launched only in a 5G network by an insider opponent who has access to the core network. In the following, we describe the four DoS-related attacks, whereas the two remaining privacy-related attacks are analyzed in Section 2.6. The first attack is the *stream reuse attack*, where the attacker impersonates a 5G Network Function (NF), causing stream ID and connection exhaustion to legitimate NFs. In the *flow control attack*, the attacker manipulates the WINDOW_UPDATE size to consume the server's resources, whereas in the *dependency and priority attack* the dependency tree storing request priorities can be exploited to consume the server's memory. Finally, in the *header compression attack* the attacker creates a compressed message that consumes a large amount of memory when decompressed.

From the above-mentioned attacks, only the stream reuse could be possible against HTTP/3. The flow control and the dependency and priority attacks cannot be exploited in HTTP/3, as stream-level multiplexing is provided by QUIC. Similarly, the header compression attack is inapplicable to HTTP/3 as the compression mechanism (HPACK) has been replaced by QPACK. Furthermore, it mainly depends on the existence of a zero-day vulnerability on the attacked endpoint, which is irrelevant of the HTTP version used.

The authors of Ling et al. (2018) proposed the H$_2$DoS attack, a novel application-layer DoS attack against HTTP/2 that exploits its multiplexing and flow control mechanisms. Specifically, they capitalized on two different frame types of HTTP/2 that play an important role in flow control, namely, SETTINGS and WINDOW_UPDATE. Similarly to previous attacks, H$_2$DoS exhausts server resources by initiating and maintaining active a huge number of HTTP/2 connections. The authors demonstrated that their attack could lead to a DoS by occupying all available connections; they also compared their attack against two other well-known DoS assaults, namely, *slowloris* and *thc-ssl-dos*. The results showed that H$_2$DoS was more effective in comparison to the other two, raising both CPU and memory usage to ≈40% and 10%, respectively. On the other hand, their comparison showed that slowloris consumed more CPU after 12 min, having an average of 50% CPU usage, whereas H$_2$DoS dropped to 30%. Regarding the repeatability of this attack, it should be noted that not all the necessary information, such as field values, are available. Again, as with other similar attacks, H$_2$DoS is infeasible in HTTP/3, as the above-mentioned flow control fields were removed.

The work in Praseed and Thilagam (2020) proposed a new DDoS attack, dubbed Multiplexed Asymmetric attack, where computationally intensive requests are multiplexed together. The main scenario tested by the authors was sending multiple requests to cause CPU exhaustion to the HTTP server. On top of this application layer attack, if the server supported Server Push, a flooding DDoS attack was triggered at the network layer. The Server Push feature used in this last case is responsible to preemptively deliver data packets to the client before even requesting them. Both HTTP/1.1 and HTTP/2 servers were tested against the Multiplexed Asymmetric attack under the same load, and the results showed that the HTTP/2 version was more resilient. Also in this case, the necessary information to reproduce the attack, such as the attack scripts, are not available. Regarding the migration of the attack to the latest HTTP version, while HTTP/3 supports multiplexing and Server Push, they are implemented with different mechanisms, making these attacks not directly applicable.

### 2.4. Slow rate attacks

Even though slow rate attacks are essentially a subcategory of DoS attacks, we chose to present them separately due to their stealthier nature that requires more effort and different detection methods. In Adi et al. (2015), a DoS attack variant was introduced, which is based on sending low-rate traffic that contains resource-hungry instructions to a victim HTTP/2 server. This work takes advantage of the same flow control vulnerability that manipulates the WINDOW_UPDATE size as in Adi et al. (2016), which has been analyzed in Section 2.3. The authors, using a custom testbed, answer three main questions: (a) how DoS attacks towards an HTTP/2-enabled server can be mounted, (b) how many servers can a single client instance attack successfully, and (c) how can attacks be stealthier by introducing time delays in the attack traffic. The experimental evaluation involved five different test cases, all of which had an increased CPU usage between 88% and 98%, showing that a DoS attack is feasible. Regarding (b), an attacker with only a single client was able to assault successfully 12 server machines and this shows that it is possible to interrupt a large number of HTTP/2 services even without a DDoS attack.

Finally considering (c), the introduction of time delays from 1 to 100 ns did not make the attack stealthier, suggesting that slow rate attacks against HTTP/2 are impracticable. Nevertheless, we argue that the delay of 1 to 100 ns is too short to consider the attack a slow rate one. For instance, the analysis in Shorey et al. (2018) demonstrated that a slowloris assault needed ≈5,882 packets per second on an HTTP connection, whereas in the current attack with the lowest-rate scenario (one packet every 100ns) 10 million packets were sent in the same duration, resulting in 1,700 times more packets. In any case, given that the WINDOW_UPDATE has been removed in HTTP/3, this attack is not directly applicable in the latter.

The authors of Zhang and Shi (2018) proposed zAttack, a new slow rate DoS attack that exploits the invalid frame state vulnerability of HTTP/2 by manipulating the SETTINGS_MAX_CONCURRENT_STREAMS field. The target is to establish a large number of open streams that render the server unable to serve additional requests, resulting in a DoS.

The authors tested their attack against three different web servers, namely, Apache2 v2.4.33, Nginx v1.14.0, and H2O v2.3.0. The results show that each server had a different timeout period, i.e., 60, 300, and 10 secs, for Apache2, Nginx, and H2O, respectively. It was also observed that the maximum number of simultaneous connections that each server could handle was 400, 1024 and 2030, respectively. The required rates to bring down the servers are 6.7, 3.4, and 203 requests/sec; these data suggest that in the H2O case the attack could be more easily detected. Regarding HTTP/3, it is not possible to mount the zAttack given that the SETTINGS_MAX_CONCURRENT_STREAMS field was removed in the latest HTTP version Bishop (2022).

The authors of Tripathi and Hubballi (2018) examined slow rate DoS attacks in HTTP/2 and proposed an anomaly-based detection method. Specifically, they presented how slow rate DoS at-tacks can consume a web server's connection pool by injecting specially crafted HTTP requests. Then, they tested their proposals using popular web servers, namely, Nginx 1.10.1, Apache 2.4.23, Nghttp2 1.14.0, and H2O 2.0.4, in their default settings and the results showed that most of them are vulnerable to the proposed assaults.

Next, they implemented five different attacks that consume the server's resources by making it wait for data that never arrive. The first attack exploits the WINDOW_UPDATE and the authors found that Nginx and Nghttp2 waited for 60 sec, Apache for 300 sec, and H2O waited indefinitely. This attack is infeasible on HTTP/3, given that the WINDOW_UPDATE field was removed from the specification. Two of the attacks manipulate the HEADERS frame; in the first one, Nghttp2 waited for a maximum of 975 sec, while Apache, Nginx, and H2O waited indefinitely on repeated attacks. In the other attack, Apache and H2O waited indefinitely, while Nginx and Nghttp2 waited for 90 and 60 sec, respectively. Regarding both of these attacks, RFC 9114 Bishop (2022) defines that HTTP/3 does not provide an explicit priority assignment mechanism (such as HEADERS), rendering this attack not directly applicable. Another attack exploited the Connection Preface message and Nginx waited indefinitely on repeated attacks, while Apache, H2O, and Nghttp2 waited for 300, 10, and 975 sec, respectively. This attack is not possible on HTTP/3 since the Connection Preface message is not part of the specification. Finally, the last attack manipulates the SETTINGS frame, leading Apache, Nginx, H2O, and Nghttp2 to wait for 5, 180, 10, and 975 sec, respectively. Given that HTTP/3 still implements control frames, i.e., SETTINGS, it is possible that this attack is still applicable to HTTP/3.

As a defensive measure, the authors proposed an anomaly-based technique to detect these types of attacks, which works by comparing observed traffic with expected patterns. Their method was able to detect these attacks with high accuracy.

*2.5. HTTP/2 smuggling attacks*

In Jake Miller (bishopfox) (2022), the port of HTTP request smuggling to HTTP/2 is investigated. The author exploited the Upgrade header (101) to upgrade HTTP/1.1 connections to HTTP/2 over cleartext (h2c), while having a reverse proxy as an intermediate. The result of such an attack is that a malicious client can establish unrestricted HTTP connections with back-end servers. This way an attacker is able to bypass reverse proxy access controls or restrictions such as accessing a directory. Although this attack is considered a misconfiguration, the author suggests blocking Upgrade requests or limit them only to the necessary services (e.g., websocket). Given that HTTP/3 does not have a cleartext mode, this attack does not apply to it.

The work in James Kettle (2022) illustrated different techniques against web applications to create an HTTP request smuggling attack, the Achilles heel of HTTP/1.1 protocol. While the issues mentioned were patched by the respective website owners, similar techniques can possibly affect other web implementations due to different HTTP/2 misconfigurations. The author presented the following three HTTP/2 desync scenarios, in which an HTTP request smuggling attack was feasible through an HTTP/2 connection:

1. HTTP/2 desync attack: Such an attack can occur when a front-end server communicates with clients on HTTP/2, but uses HTTP/1.1 to communicate with the back-end server. The main cause of such attacks is that the front- and back-end cannot agree on which of the Content-Length or Transfer-Encoding header to use for obtaining the request length. This type of attacks can allow an attacker to inject arbitrary prefixes to HTTP requests of other users, steal passwords and credit card numbers, or even make the front-end send the response intended for a user to a different user.
2. Desync-powered request tunnelling: This is a subclass of the previous attack, and it relies on the connection-reuse strategy followed by the front-end. When a request arrives at the front-end, it has to decide whether it will forward it using an already established connection with the back-end or create a new one; this decision affects the possible attacks that can be mounted. The range of assaults includes requests reaching the back-end without being processed by the front-end, exploiting internal headers injected by the front-end, and web cache poisoning.
3. HTTP/2 exploit primitives: Different exploit techniques were illustrated against HTTP/2 in this case. For instance, it is possible to send requests with multiple methods or paths, lead to server-side request forgery (SSRF), enable request line injection which allows bypassing block rules in the back-end server, and tamper with internal and external headers.

While these assaults were identified mostly against web applications, it is possible that they could also be used against other HTTP/2-based connections. Furthermore, even though RFC 7540 protects from such methods (for example, the transfer encoding header field is forbidden in HTTP/2) some servers accepted it. For this reason, it is possible to affect HTTP/3-enabled servers as well.

*2.6. Privacy attacks*

Suresh et al. (2018) provided an overview of the HTTP/2 protocol and a short discussion on security issues, such as Head-of-Line Blocking and DoS attacks. The main objective of this work was to investigate the feasibility of decrypting HTTP/2 traffic, using a suitable HTTP/2 test environment. The authors found out that by exploiting the SSLKEYLOGGING feature, i.e., a mechanism used by browsers on the client side to log private keys into a file, it is possible for an attacker to extract private information, such as websites visited, Operating System (OS), and browser version. However, the authors neither provided a complete survey on existing HTTP/2 attacks nor examined the possibility of their migration to HTTP/3. Since this issue pertains to TLS decryption, it is considered pertinent to HTTP/3.

In Hu et al. (2018), a MitM attack against a 5G network using HTTP/2 for service-based interfaces is presented. In this assault, the opponent first performs a DNS poisoning attack to insert a malicious NF between two legitimate NFs. Then, similar to Kampourakis et al. (2022); Zhang et al. (2020), the attacker can intercept and read TLS-encrypted traffic data by presenting a forged TLS certificate to the communicating parties. Another family of privacy attacks that can be mounted in the same setting is interconnection attacks. Opponents can track users or eavesdrop private information when different networks interconnect if the existing security mechanisms are misconfigured or not deployed at all. To succeed in the aforementioned attacks, the attacker should have access to the 5G core network. Regarding portability to HTTP/3, MitM and interconnection attacks are irrelevant of the HTTP protocol, that is, they could be possible either with or without the existence of HTTP/2 or HTTP/3 in the absence of proper security mechanisms.

The authors of Lin et al. (2019) compared the resilience of both HTTP/1.1 and HTTP/2 against state-of-the-art web fingerprinting attacks. Specifically, they collected the 99 most popular websites as ranked by Alexa Amazon (2022) to get the dependency structure of each site as well as the size of each site's resources. The Chrome DevTools protocol was used to log requests and responses sent or received by the browser, as well as intercept network events, such as requestWillBeSent, responseReceived, dataReceived, and loadingFinished. The authors used this information to create models of the features that influence the network trace of loading a page. These models were then served on both HTTP/1.1 and HTTP/2, using the Caddy web server to compare their susceptibility to fingerprinting. Additionally, the *tcpdump* tool was used to export network traffic to pcap files. These files were then processed to filter out DNS packets, and clear out TCP packets with no data, recording only the direction, size, and timing of each packet. These attributes were used as input in a random forest model to perform fingerprinting. According to the results, the model in HTTP/1.1 achieved an accuracy of 80%-99%, while in HTTP/2 with server push enabled the accuracy diminished to 74.2%, showing a smaller attack surface. According to Smith et al. (2021), QUIC protocol can evade up to 96% of TCP-trained classifiers; however, they conclude that QUIC shows a similar difficulty of fingerprinting as TCP.

The work in Mitra et al. (2020) showed that it is possible to break the privacy offered by HTTP/2 multiplexing. HTTP/2 allows concurrent server threads to process multiple objects, resulting in multiplexed object transmission. This feature is useful for avoiding Head-of-Line blocking, i.e., a large object in the queue blocking the subsequent objects from being processed, which has been widely exploited in HTTP/1.1 to perform traffic analysis. Furthermore, HTTP/2 multiplexing makes it difficult for a passive attacker to identify individual objects over TLS traffic, and for this reason it is used as a basis for relevant privacy schemes. The authors assumed that an attacker may alter network parameters, namely latency, jitter, bandwidth, and packet drops, to introduce spacing between consecutive GET requests, which are sent to a server. This process can block the opportunity for the server to multiplex the objects corresponding to these requests, thus, negating the privacy benefits that come with it. The experimental results using the above parameters showed that:

- a uniform delay introduced for all packets is not effective for the described attack,
- the introduction of jitter so that the inter-arrival time of requests is 50ms results in 54% of objects not being multiplexed,
- a bandwidth reduction of 20% resulted in over 60% of non-multiplexed cases, and
- an 80% packet drop starting when the object of interest is sent for at least 6 sec resulted in 90% non-multiplexed cases.

As a remediation, the authors propose that some features of HTTP/2, namely server push and prioritization, can be used to set a different object priority and confuse the attacker. Regarding HTTP/3, it should be further examined if such attacks are still applicable, since the latest HTTP protocol version uses multiplexing and streams to transfer data.

*2.7. Attack taxonomy*

A taxonomy of the attacks reported in this section is presented in Fig. 1. The attacks can generally be classified into two broad categories based on their HTTP version relevance: the ones that apply only in HTTP/2 and the ones that apply in HTTP/2 but could also apply in HTTP/3. Based on their characteristics, the studied attacks are classified into five categories: amplification, cryptojacking, DoS, slow-rate DoS, smuggling and privacy. As already explained above, even though slow-rate is a special subcategory of DoS, we chose to examine it separately due to being more difficult to detect.

A first observation by counting the attacks in Fig. 1 is that their majority (around 63%) are DoS (DoS and slow-rate DoS in the taxonomy diagram). Another major remark is that about one third of the attacks (close to 38%) can be ported to HTTP/3, mainly due to the different mechanisms used for flow control. Regarding individual categories, all the privacy-related attacks can be ported to HTTP/3, showing that the different characteristics of the new protocol version do not affect privacy-intrusive methods.

## 3. Hands on evaluation

For the hands-on evaluation, we reused the contemporary testbed given in § 5.1 of Chatzoglou et al. (2022d). Precisely, this testbed is composed of the currently six most popular QUIC- and HTTP/3-enabled server implementations, namely OpenLiteSpeed, Caddy, NGINX, H2O, IIS, and Cloudflare. The reader should keep in mind that, at the time of writing, paradoxically, some servers like Algernon Algernon (2022) do endorse QUIC, without however supporting HTTP/3. In total, four attacks were tested against each server; two of them stem directly from the HTTP/2 literature, that is, flooding and slow-rate, while the rest, that is, downgrade and HTTP/3-tables/streams that are presented in Sections 3.2 and 3.4, are new. The results per attack on each server implementation are recapitulated in Table 1. Note that the identified vulnerabilities are due to the specific implementations and not the HTTP/3 protocol itself. Among others, this may be due to either a software bug, a misconfiguration, an unwitting supposition, or a misconception regarding the RFC during the software development phase. The relevant attack scripts are available at a public GitHub repository[1].

*3.1. HTTP/3 flooding attack*

For this attack, the *curl* library with HTTP/3 enabled was used; note that at the time of writing, HTTP/3 and QUIC support in curl is unripe. To this end, we relied on the Docker image provided in *curl-http3*[2] repository in GitHub. First, we built locally the docker image by using the Dockerfile of that repository. Next, through the `docker exec` command, we connected to the Docker and executed the attack.

The attack used the *bash* capabilities to utilize 10 parallel curl requests; each request was executed for 1 sec (timeout). The curl command had the *GET* method as the primary one along with three additional method headers, namely, *HEAD, POST*, and *GET*. Also, a custom "settings" header with the 0 value was included, together with 26 bytes of null data to be sent along with each HTTP request.

On Caddy, the result of this tactic is a CPU usage of 99.9% 30 sec after initiating the attack, thus, paralyzing the server. Moreover, after the attack was active for 30 sec, Cloudflare presented an increased delayed response time of more than 3 sec with each request. The remaining servers coped well with this attack, i.e., they only suffered a heightened (<5%) CPU usage.

*3.2. HTTP/x downgrade attack*

Each server has been specifically setup to only communicate with the HTTP/3 protocol. However, it was observed that if TCP data were allowed to pass the firewall, the server responded to HTTP/1.1 requests establishing an HTTP/1.1 connection. Interestingly, the server admin is provided with no option to disable HTTP/1.1, but only to block the TCP protocol via the firewall. In this respect, if the firewall allows TCP traffic, the attacker may be able to mount HTTP/1.1 protocol relevant attacks, such as, HTTP request smuggling ones Chatzoglou et al. (2022a); James Kettle (2022).

Even worse, in addition to HTTP/1.1 connections, three out of the six servers, namely H2O, IIS 10, and Caddy, allowed HTTP/2 connections (without being configured as such), thus further increasing the server's attack surface. It can be argued that, for the sake of backwards compatibility, enabling by default HTTP/x protocols is desirable. However, this capability should be offered to the server admin in an opt-in/opt-out basis, which is not the case for the affected servers.

---

[1] https://github.com/efchatz/HTTP3-attacks.
[2] https://github.com/unasuke/curl-http3.

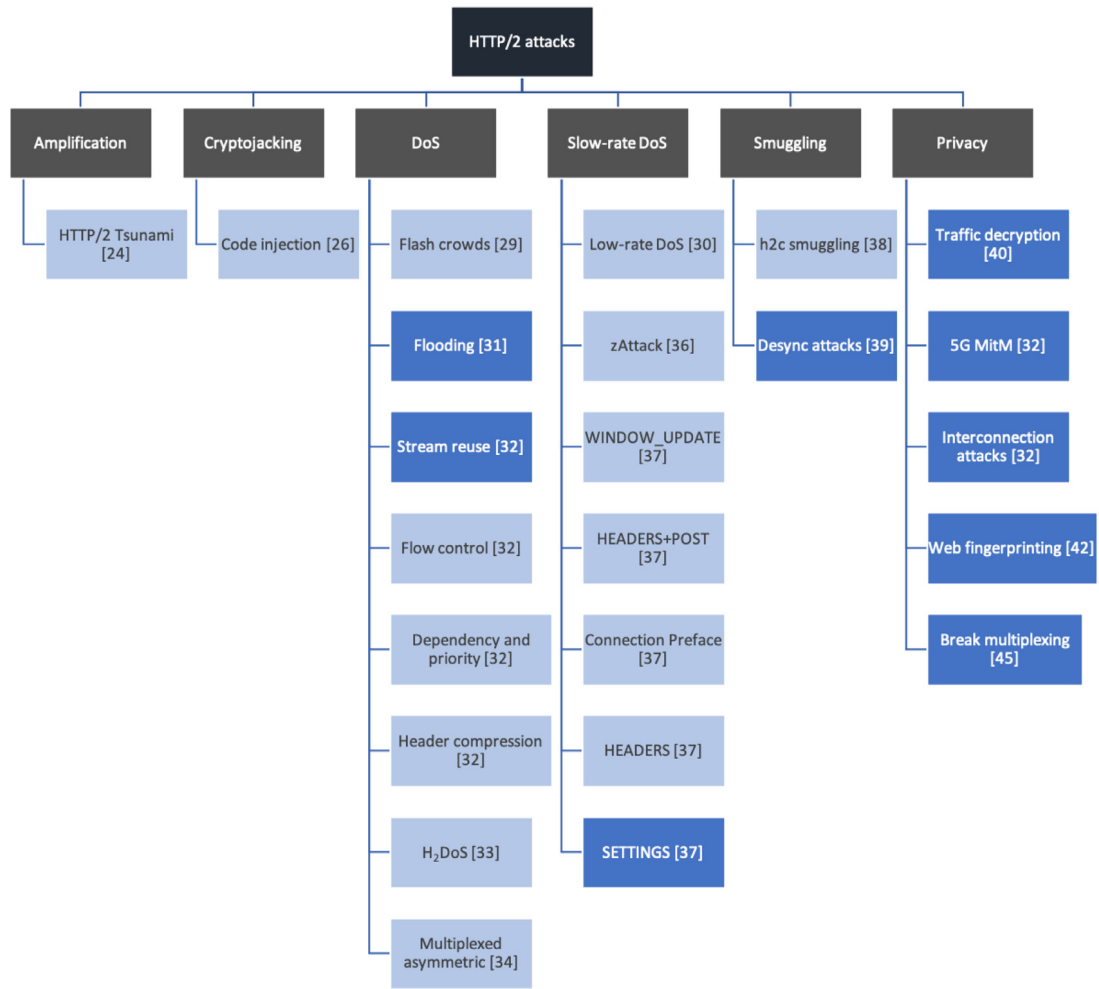

**Fig. 1.** Taxonomy of attacks against HTTP/2 (light blue: attacks against HTTP/2 only, blue: attacks against HTTP/2 that apply also to HTTP/3).

**Table 1**
Vulnerabilities identified in each server.

| Name | OpenLiteSpeed | Caddy | NGINX | H2O | IIS | Cloudflare | Total |
|---|---|---|---|---|---|---|---|
| HTTP/3 flooding | ✗ | √ | ✗ | ✗ | ✗ | √ | 2 |
| HTTP/x downgrade | √ | √ | √ | √ | √ | √ | 6 |
| Slow-rate HTTP/3 POST | ✗ | √ | ✗ | ✗ | ✗ | ✗ | 1 |
| HTTP/3-tables/streams | √/✗ | √/√ | √/√ | √/√ | √/✗ | ✗/√ | 6 |
| Total issues | 2 | 5 | 3 | 3 | 2 | 3 | – |

### 3.3. Slow-rate HTTP/3 POST attack

Another slow-rate type of attack was tested against all the servers, this time using a different HTTP method, namely the *POST* one. The attack script initiates about 40 parallel connections, with each one terminated after 5 sec. For this attack, we also employed the *OpenSSL* library for generating custom and random payloads of 32 bytes, which were sent to the targeted server. Caddy was the only server affected by this attack variation; the server's CPU usage was increased, thus delaying its responses to the clients trying to fetch a webpage.

### 3.4. HTTP-tables/streams attack

The current attack tampers with the values of *max_table_capacity* and *blocked_streams* parameters. These two parameters were introduced with the new control HTTP/3 frames, and they are transmitted with the so-called *SETTINGS* frame as the last fields related to HTTP/2. Note that the default values for these two parameters are 4096 bytes and 16 streams, respectively. We experimented with both small and large values for these fields, namely, 16 bytes and 4 streams and 409,600 bytes and 1,600 streams, respectively. By exploiting the *aioquic* Python library, we created 100 parallel connections to the targeted server with a timeout of 5 sec. This means that some connections were dropped before their completion.

Each attack lasted for about 2 min. Regarding the parameters' small values, IIS 10 and H2O presented a significant delay of around 3 sec in their HTTP responses. On the other hand, OpenLiteSpeed and Nginx paralyzed, being unresponsive for about 10 to 30 sec. Cloudflare seems to be largely immune to this attack, nevertheless, all the servers but Caddy suffered an increased CPU usage between 10% and 15% while the attack was ongoing. Even worse, the CPU usage in the Caddy server reached 99.9% just after the first sec of the attack.

Similar observations were made when testing higher values for these two fields. Cloudflare presented a delay that exceeded 3 sec, but only for new connections. H2O suffered an additional response time of more than 4 sec, Caddy showed a CPU usage of 99.9%, responding with an excessive delay to new and existing connections, and Nginx was paralyzed, being unresponsive for about 10 to 30 sec. It can be assumed that these issues are related to HTTP/3 (or even QUIC) libraries used by each server, and they are a clear indication that new implementations need further examination before their deployment in real-life environments.

Given the severity of this attack, following a Coordinated Vulnerability Disclosure (CVD) process, we informed the affected vendors about the underlying vulnerability. To track this issue, MITRE assigned CVE-2022-30592, which received a base Score of 9.8 (critical)[3]. At the time of writing, only LiteSpeed has released a patch in *lsquic*[4] to mitigate this issue, which basically triggers a Null Pointer Dereference. Precisely, the fix comes in the form of zeroing the value of any *max_table_capacity* parameter that is lower than 32.

## 4. Dataset

As already pointed out, in the context of this work, and in view of the review presented in Section 2, the results presented in Section 3 and in § 5.2 of Chatzoglou et al. (2022d), we create the first to our knowledge dataset considering attacks on HTTP/2, HTTP/3, and QUIC.

A preliminary evaluation of the dataset by means of legacy machine learning methods is also offered. We anticipate that the publicly provided dataset[5] along with its evaluation will serve as a common basis and guidance for future work.

The dataset, dubbed "H23Q" comprises a total of 10 assaults:

- Those given in Table 1, but the HTTP/x downgrade one.
- The *quic-flooding, quic-encapsulation, quic-loris*, and *quic-fuzz* assaults introduced in § 5.2 of Chatzoglou et al. (2022d).
- An HTTP request smuggling attack plus two traditional attacks from the HTTP/2 domain: (i) A flooding one, which sets a hefty *max_concurrent_request* value equal to 100K, and (ii) a pause-resume flooding, which repeatedly creates HTTP/2 connections; the connections are paused and then resumed. The latter two attacks were included for the sake of completeness, since no work so far offer an HTTP/2 security-oriented dataset. It should be noted that the HTTP-request smuggling assault has a similar effect to the HTTP/x downgrade one.

The H23Q dataset is offered in both pcap and CSV formats. Precisely, the CSV files are labelled and contain 200 features, i.e., 199 generic ones and the label class. We also include several Python scripts and instructions on how to generate new CSV files with additional set of features and how to label them.

### 4.1. Testbed

The testbed for the creation of the dataset comprised six different HTTP/3-enabled servers, which run on the Azure cloud infrastructure. The hardware specifications of all the employed machines, servers and clients, are summarized in Table 2. The utilized clients were operated from three different subnetworks. The first comprised six clients, the second three, and the last four, with one of them operated by the attacker. Each client and server received its last update on April 30, 2022.

To replicate real-life traffic scenarios, two of the public network interfaces (DSL routers), changed their public (routable) IP address during the recording process. This means that some attacks contain different public IP addresses for the same clients. Regarding the configuration of each deployed server, the interested reader is referred to § 5.1 of Chatzoglou et al. (2022d). Note that IIS 10, H2O, and Caddy enable HTTP/2 by default.

Fig. 2 depicts a high-level view of the network topology. The red-colored client in the third subnetwork represents the attacker, while the orange-colored ones are part of the botnet the attacker created for the needs of specific attacks. Each server was behind a DNS zone. The latter was assigned with a registered domain name, and then, each server was assigned with a unique subdomain. Each HTTP server offered a simple HTML webpage. For some indefinite reason, some servers, including IIS 10, experienced problems in offering every time its webpage over HTTP/3. So, these servers communicated with the clients with either HTTP/2, if they supported it by default, or HTTP/1.1. As a result, in the dataset, apart from the HTTP/3 normal traffic, one can observe HTTP/2 and HTTP/1.1 normal traffic as well.

Client-server communication was done based on a random pattern. Precisely, the *Selenium* Python library was installed in each client, and each one of them requested randomly to connect to an HTTP server. Then, the client waited for 5 sec, and retried to communicate with another or the same server after a random sleep time ranging between 1 and 5 sec. The HTTP connection was made either through the Chrome or the Firefox browser. To enable the decryption of the recorded traffic, all the clients, including the attacker's one, stored their TLS keys locally.

Given that the H23Q traffic was generated to a greater or lesser extent evenly from 13 client devices sending queries pseudorandomly with intervals of 1 to 5 sec, the dataset approximates real-life network traffic, which often encompasses a much greater number of clients connecting to a web server concurrently, with assorted peaks and drops in demand. On the other hand, for obvious reasons, organizations are highly reluctant in granting their datasets, and this especially applies to network traffic which does contain cyberattacks. Even in case such a dataset is offered, certain fields will be probably anonymized, encrypted (TLS traffic for instance), or stripped off for understandable reasons; this may be a perplexing factor when applying machine learning methods to the dataset. And naturally, this scarcity of publicly available real-life datasets is especially evident for new technologies and protocols as the QUIC one. In any case, the H23Q dataset is labeled, with the option to have the TLS traffic in each pcap file decrypted (the TLS decryption keys are provided along with the dataset), therefore, anyone can separate between the two traffic classes and potentially merge it with any matching network traffic corpus containing benign or attack records.

### 4.2. Data collection

Regarding the data collection, the following points are important:

- The network traffic capturing process was performed on each server separately. The reason behind this choice is that by recording traffic in this way, it offers more flexibility in terms of a possible IDS, either a network-based or host-based IDS.

---

[3] https://nvd.nist.gov/vuln/detail/CVE-2022-30592.
[4] https://github.com/litespeedtech/lsquic/releases/tag/v3.1.0.
[5] The "H23Q" dataset is available for download at the well-known AWID website at https://icsdweb.aegean.gr/awid/other-datasets/H23Q.

**Table 2**
Specifications of servers and clients in the testbed. For clients 1 to 9, a different but unique routable IP address may exist in each pcap file. For instance, the "http-flood" pcap file contains only the 5.203.250.215 public IP address associated with these clients.

| Name | OS | CPU/RAM | Version | Network | IP address (router) |
|---|---|---|---|---|---|
| OpenLiteSpeed | Ubuntu 18.04 | 1/1 | 1.7.15 | Azure | 10.0.0.4 |
| Caddy | Ubuntu 18.04 | 1/1 | 2.4.6 | | 10.0.0.5 |
| NGINX | Ubuntu 18.04 | 2/4 | 1.21.7 | | 10.2.0.4 |
| H2O | Ubuntu 18.04 | 1/1 | 2.3.0-DEV | | 10.0.0.4 |
| IIS | Windows Server 2022 | 2/4 | 10 | | 10.0.0.5 |
| Cloudflare | Ubuntu 18.04 | 2/4 | 1.16.1 | | 10.1.0.4 |
| Client 1 | Windows 11 | 2/4 | 21H2 | I | 5.203.250.215 or |
| Client 2 | Windows 11 | 2/4 | 21H2 | | 5.203.228.219 or |
| Client 3 | Windows 10 | 2/4 | 21H2 | | 5.203.253.63 |
| Client 4 | Windows 10 | 2/4 | 21H2 | | |
| Client 5 | Ubuntu 22.04 | 2/4 | 5.15.0-27 | | |
| Client 6 | Ubuntu 22.04 | 2/4 | 5.15.0-27 | | |
| Client 7 | Windows 11 | 2/4 | 21H2 | II | 5.203.165.59 or |
| Client 8 | Windows 10 | 2/4 | 21H2 | | 5.203.215.191 or |
| Client 9 | Windows 10 | 2/4 | 21H2 | | 5.203.255.68 |
| Client 10 | Windows 10 | 2/4 | 21H2 | III | 85.75.109.194 |
| Client 11 | Windows 10 | 2/4 | 21H2 | | |
| Client 12 | Ubuntu 22.04 | 2/4 | 5.15.0-27 | | |
| Attacker | Ubuntu 20.04 | 4/16 | 5.13.0-40 | | |

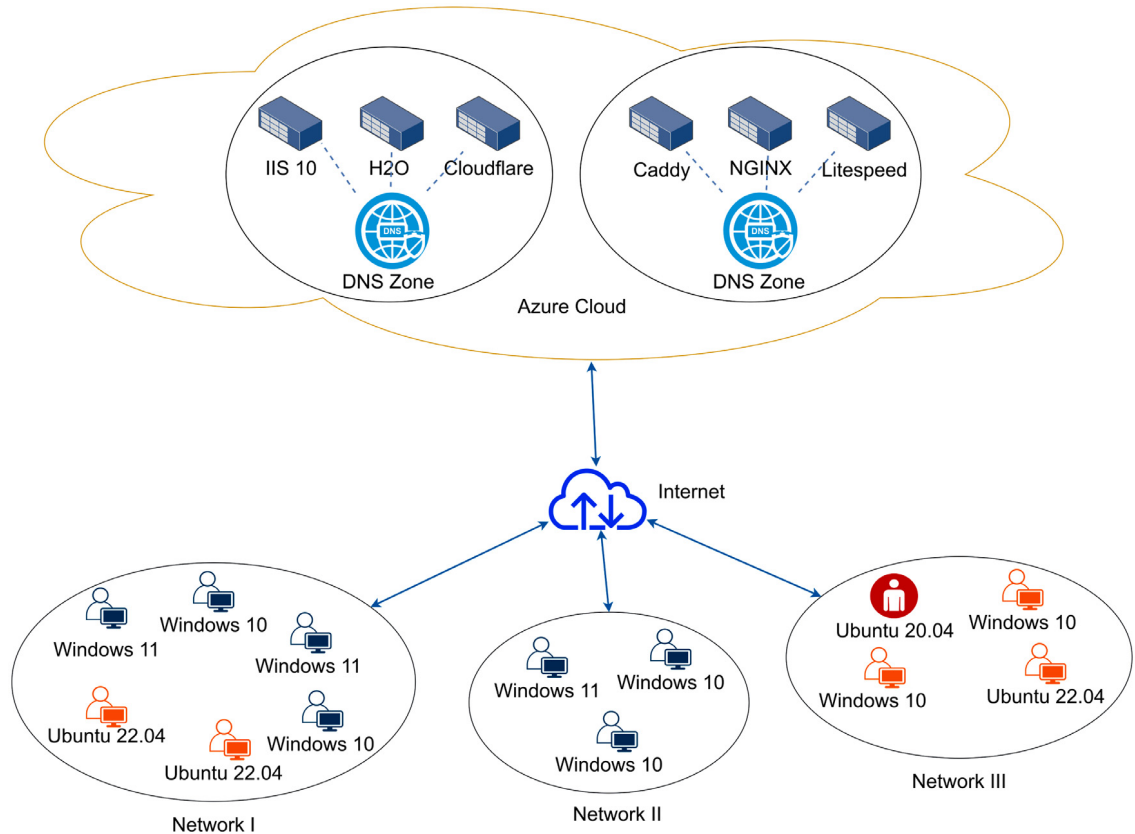

**Fig. 2.** Network topology used in the creation of the dataset. The red node refers to the attacker, while the orange ones to the devices that belong to the attacker's botnet.

- To this end, each attack was split into six different pcap files, one per server. And since the dataset contains 10 attacks, the dataset comprises 60 pcap files.
- To reduce the size of the dataset, we recorded around 1M packets per attack, meaning that each server captured approximately 150K packets.
- As mentioned in Section 4.1, each client stored their TLS keys. This enables the decryption of the corresponding traffic in the dataset.

- The *Wireshark* v3.6.3 utility was installed on each server for capturing the incoming and outgoing traffic. For Ubuntu-based servers, *tshark* was used; the latter utilizes Wireshark to capture the traffic. All the captured processes were filtered by appropriate IP and port filters for not recording any unwanted traffic.
- The attack parameters, including its duration, the frames per second rate, and the use of bots or not, differ depending on the attack type. The purpose was to trace out each attack the best way possible.

## 4.3. Attacks in the dataset

As already mentioned, the dataset comprises 60 pcap files, that is, 10 attacks × 6 servers. Each attack against a server was performed following the same order, i.e., OpenLiteSpeed, Caddy, NGINX, IIS, Cloudflare, and H2O. We detail each attack below, while Table 3 recaps the characteristics of each attack as seen in the corresponding files of the dataset.

- *HTTP3-flood*: Multiple HTTP/3 requests were sent to each server. To achieve this, the attacker utilized *curl* v7.83.0 along with the local bots for creating a DDoS effect. The total time of this assault was 10 min, with the first 4 being normal traffic. After that, each server was under attack for 1 min.
- *Fuzzing*: It contains fuzzing traffic and a number of packets similar to those of a *hash-collision* attack Preneel (2011). The opponent attacked solo, without the use of bots. Again, the assault lasted for 10 min, with the first 4 being normal traffic and the remaining 6 devoted to attacking each server for 1 min. The attacker utilized the *Fuzzotron* fuzzer along with the *Scapy* Python library for crafting custom packets. This assault should be considered as a transport layer one, since most of the attack traffic consists of UDP datagrams.
- *HTTP3-loris*: This assault has the same duration as the previous two. The basic difference here is that the attacker utilized its local and remote bots as well, thus, causing a considerable DDoS effect. Recall that, with reference to Fig. 2, the attacker's botnet comprises bots residing in networks I and II. The local bots issued simple HTTP requests, while the remote bots and the attacker were placing an HTTP request with a random payload every 5 sec; the data were generated through the *OpenSSL* library.
- *HTTP3-stream*: This attack was done in two cycles, carried out back to back. The first, follows the same timing scheme as the previous three assaults: the first 4 min for normal traffic and the rest 6 for the attack. In this cycle, the attacker utilized the *aioquic* Python library v0.9.20, and as mentioned in Section 3.4, they increased by far the *max_table_capacity* and *blocked_streams* values. The second cycle comprises 3 min of normal traffic, and after that, the attacker mounted a variation of the assault where the aforementioned two fields had a low value. The total duration of this attack is 20 min, with the attacks laying from the 4th to the 10th, and from the 13th to the 19th min.
- *QUIC-flood*: With the aid of the *aioquic* library, the attacker instructed the local bots to perform a QUIC-flood. The timing scheme of the current attack is identical to the first three ones.
- *QUIC-loris*: The attacker exploited all its bots as well, therefore, increasing the DDoS effect. The connection requests crafted with the help of the *aioquic* library were placed every 5 sec from the attacker and both the local and remote bots. The attack phase is between the 4th and the 10th minutes.
- *QUIC-enc:* The methodology of the current assault is similar to the *quic-encapsulation* one detailed in Chatzoglou et al. (2022d). Through the *Scapy* library, the attacker sends custom encapsulated packets, which are in the form of *IP(UDP(IP(TCP)))* and *IP(UDP(IP(UDP)))*. The timing scheme of the current attack is identical to the first three ones.
- *HTTP-smuggling*: It has a longer duration, namely, 15 min. The aggressor initiated the attack at the 3rd min and assaulted persistently each server for 2 min. For this attack, the attacker utilized *curl* along with *OpenSSL* for generating custom payloads per packet. A different packet structure was used when attacking each server.
- *HTTP/2-concurrent*: This penultimate attack pertains to HTTP/2. Its total duration was 6 min, with the attacker launching it after the 3rd min, and changing the targeted server every 30 sec. Once more, the *curl* tool was used. Specifically, for stressing each server, the tool was instructed to create 100K *MAX_TOTAL_CONNECTIONS* with 100K *MAX_CONCURRENT_STREAMS* each; recall that the first variable defines the maximum number of simultaneous open connections of a client, while the second, the maximum count of simultaneous streams to support over a single HTTP connection. In case the target server did not enable HTTP/2, the traffic was over HTTP/1.1, thus resulting to an HTTP/1.1 flooding.
- *HTTP/2-pause*: This last assault has the same timing scheme as the previous one. Through the *curl* tool, the assailant ceases and starts the HTTP/2 threads of each connection in an attempt to paralyze the server. If the target server did not offer HTTP/2, the attack takes the form of an HTTP/1.1 flooding.

## 4.4. Signature of attacks

This section offers symptomatic signatures (footprints) of selected attacks of the dataset based on packet per second (PPS) units of measurement. Specifically, we chose four representative assaults, i.e., two HTTP/3 and two QUIC oriented. The former were taken from a specific server, while the latter depict the traffic from all the servers.

First off, Fig. 3 depicts the normal versus HTTP3-flooding traffic, but only for the Cloudflare server. As it can be observed, the attack is clearly differentiated against the normal traffic. Second, Fig. 4 footprints an HTTP3-loris assault exercised against the OpenLiteSpeed server. Such "under the radar" assaults are typically used with the purpose of bypassing certain network perimeter protection mechanisms. Indeed, as observed from the figure, the attack pattern is almost identical to that of the normal traffic. So, identifying such an attack, especially on a single server, is quite challenging.

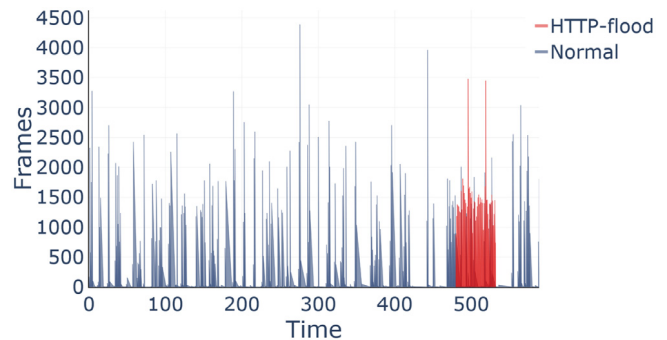

**Fig. 3.** HTTP3-flood footprint on the Cloudflare server.

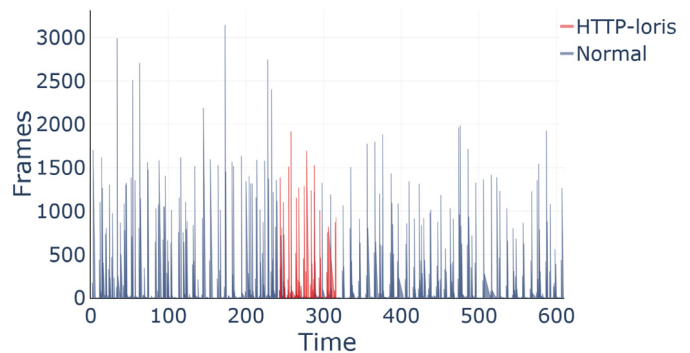

**Fig. 4.** HTTP3-loris footprint on the OpenLiteSpeed server.

Table 3

Details per attack per server. A separate pcap and CSV file is given per attack per server. An asterisk means that this attack produced a DDoS effect. The 3rd column designates the analogy of malicious to normal traffic, whereas the 7th column the malicious to total traffic. The two "traffic" columns and the "Normal/Malicious" one are expressed in number of packets.

| Per attack | | | Per server | | | | | |
|---|---|---|---|---|---|---|---|---|
| Attack name | Normal/Malicious | % | Server | Total traffic | Malicious traffic | % | Size (MB) | Duration |
| HTTP3-flood* | 1,316,770/498,810 | 37.88 | LiteSpeed | 329,264 | 112,466 | 34.15 | 281 | 6/10 |
| | | | Caddy | 142,218 | 64,082 | 45.05 | 125 | |
| | | | NGINX | 175,211 | 71,588 | 40.85 | 164 | |
| | | | IIS | 146,020 | 87,720 | 60.07 | 131 | |
| | | | Cloudflare | 342,777 | 76,747 | 22.38 | 319 | |
| | | | H2O | 181,278 | 86,207 | 47.55 | 195 | |
| Fuzzing | 660,412/22,224 | 3.36 | LiteSpeed | 191,456 | 2,440 | 1.27 | 157 | 6/10 |
| | | | Caddy | 47,703 | 3,085 | 6.46 | 55 | |
| | | | NGINX | 79,849 | 3,175 | 3.97 | 76 | |
| | | | IIS | 55,666 | 6,799 | 12.21 | 61 | |
| | | | Cloudflare | 227,949 | 2,552 | 1.11 | 225 | |
| | | | H2O | 57,787 | 4,173 | 7.22 | 75 | |
| HTTP3-loris* | 677,240/74,572 | 11.01 | LiteSpeed | 217,093 | 24,060 | 11.08 | 180 | 6/10 |
| | | | Caddy | 77,846 | 14,579 | 18.72 | 79 | |
| | | | NGINX | 79,436 | 7,210 | 9.07 | 73 | |
| | | | IIS | 56,631 | 9,566 | 16.89 | 59 | |
| | | | Cloudflare | 184,311 | 15,713 | 8.52 | 177 | |
| | | | H2O | 61,922 | 3,444 | 5.56 | 75 | |
| HTTP3-stream | 1,318,226/1,063 | 0.08 | LiteSpeed | 310,957 | 436 | 0.14 | 286 | 12/20 |
| | | | Caddy | 125,561 | 52 | 0.04 | 144 | |
| | | | NGINX | 186,542 | 27 | 0.01 | 176 | |
| | | | IIS | 141,770 | 453 | 0.31 | 153 | |
| | | | Cloudflare | 381,959 | 47 | 0.01 | 374 | |
| | | | H2O | 171,435 | 48 | 0.02 | 213 | |
| QUIC-flood* | 746,608/61,340 | 8.21 | LiteSpeed | 199,894 | 8,215 | 4.10 | 184 | 6/10 |
| | | | Caddy | 72,366 | 7,798 | 10.77 | 80 | |
| | | | NGINX | 105,826 | 8,345 | 7.88 | 101 | |
| | | | IIS | 66,448 | 23,210 | 34.92 | 71 | |
| | | | Cloudflare | 206,436 | 6,550 | 3.17 | 202 | |
| | | | H2O | 95,637 | 7,222 | 7.55 | 120 | |
| QUIC-loris* | 653,451/24,269 | 3.71 | LiteSpeed | 225,238 | 15,075 | 6.69 | 193 | 6/10 |
| | | | Caddy | 55,237 | 51 | 0.09 | 61 | |
| | | | NGINX | 73,417 | 759 | 1.03 | 71 | |
| | | | IIS | 45,158 | 1,699 | 3.76 | 54 | |
| | | | Cloudflare | 188,099 | 6,632 | 3.52 | 186 | |
| | | | H2O | 66,301 | 53 | 0.07 | 85 | |
| QUIC-enc | 741,467/5,829 | 0.78 | LiteSpeed | 224,651 | 689 | 0.30 | 191 | 6/10 |
| | | | Caddy | 66,115 | 847 | 1.28 | 76 | |
| | | | NGINX | 108,556 | 1,300 | 1.19 | 104 | |
| | | | IIS | 49,456 | 1,091 | 2.20 | 58 | |
| | | | Cloudflare | 221,603 | 1,006 | 0.45 | 217 | |
| | | | H2O | 71,085 | 896 | 1.26 | 90 | |
| HTTP-smuggle | 1,331,394/3,337 | 0.25 | LiteSpeed | 252,837 | 749 | 0.29 | 247 | 12/15 |
| | | | Caddy | 122,690 | 536 | 0.43 | 139 | |
| | | | NGINX | 180,606 | 342 | 0.18 | 173 | |
| | | | IIS | 83,507 | 540 | 0.64 | 90 | |
| | | | Cloudflare | 381,765 | 403 | 0.10 | 378 | |
| | | | H2O | 129,382 | 425 | 0.32 | 168 | |
| HTTP2-concurrent | 982,768/60,273 | 6.13 | LiteSpeed | 188,046 | 3,997 | 2.12 | 134 | 3/6 |
| | | | Caddy | 211,210 | 30,694 | 14.53 | 173 | |
| | | | NGINX | 225,822 | 3,437 | 1.52 | 209 | |
| | | | IIS | 127,480 | 21,567 | 16.91 | 73 | |
| | | | Cloudflare | 184,388 | 175 | 0.09 | 181 | |
| | | | H2O | 45,821 | 403 | 0.87 | 63 | |
| HTTP2-pause | 1,141,326/53,549 | 4.69 | LiteSpeed | 339,280 | 3,043 | 0.89 | 342 | 3/6 |
| | | | Caddy | 235,461 | 23,827 | 10.11 | 191 | |
| | | | NGINX | 238,000 | 2,880 | 1.21 | 225 | |
| | | | IIS | 138,203 | 23,466 | 16.97 | 148 | |
| | | | Cloudflare | 142,630 | 227 | 0.15 | 143 | |
| | | | H2O | 47,751 | 109 | 0.22 | 63 | |

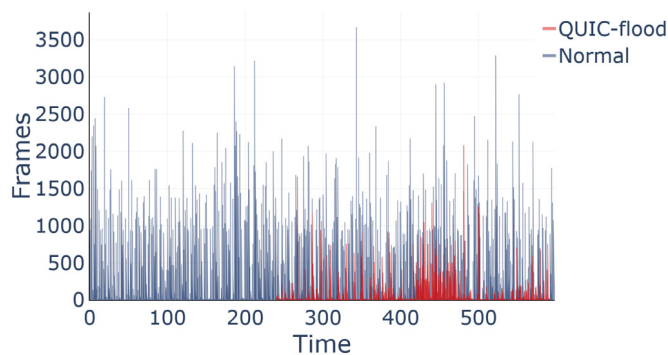

**Fig. 5.** QUIC-flood footprint on all servers.

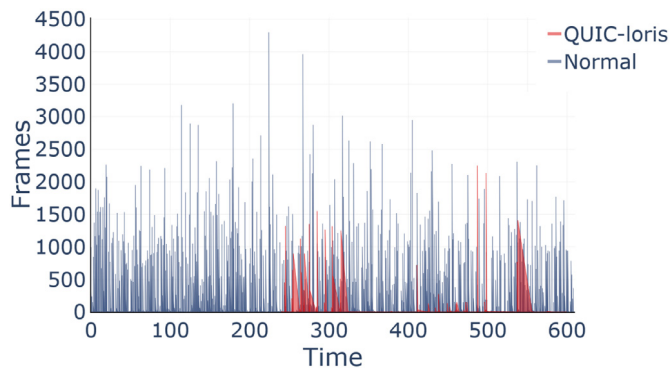

**Fig. 6.** QUIC-loris signature on all servers.

Third, Fig. 5 illustrates the QUIC-flood assault done against the six servers. A higher number of packets were captured between the 400 and 500 sec, possibly targeting the IIS server. As with the HTTP3-flood one, the attack pattern is quite easily distinguishable if compared to that of the normal traffic. Last but not least, the QUIC-loris assault is illustrated in Fig. 6. The footprint of this attack seems to be clearer in comparison to that of the HTTP3-loris, possibly due to the combination of traffic stemming from all the six servers.

## 5. Dataset evaluation

In this section, we perform an initial evaluation of the H23Q dataset through machine learning techniques, both shallow and deep learning. First, we detail the feature selection and data preprocessing procedures, and then elaborate on the experiments and the derived results. Nevertheless, similar to other well-known works introducing datasets Kolias et al. (2016), it is not in the scope of the current study to exhaustively evaluate H23Q, examining also, e.g., feature importance or conducting hyperparameter analysis, say, through the *Optuna* framework. This is left for future work.

The experiments were performed on an MS Windows 10 Pro machine with AMD Ryzen 7 2700 CPU and 64 GB RAM. For shallow classifiers, we only rely on the CPU; no GPU was utilized. We employed the *sklearn* v.1.0.1 in Python v3.8.10, for all classifiers and metrics, except for *LightGBM*. The latter algorithm was implemented with the homonymous Python library in v.3.3.2.

### 5.1. Feature selection and data preprocessing.

The following points are important regarding feature selection and data preprocessing.

- First, a large number of features (199) were extracted from the pcap files. This was a provisional set of features that could possibly assist in the identification of malicious traffic. To extract these features, we utilized *tshark*. However, before running *tshark*, the TLS keys of each client were loaded from *Wireshark*. Then, via *tshark*, we extracted the decrypted traffic of each pcap file, finally getting the initial set of the 199 features. From this large set of features, we cherry-picked less than half of them, i.e., a total of 46 features. The feature selection process was based on the study of previous work summarized in § 3.1 of Chatzoglou et al. (2022c).
- In a next step, the labelling process added one more feature to designate the attack class. Note that the Azure Cloud anonymizes the MAC addresses of the incoming traffic, therefore, every MAC address in each pcap file is anonymized as "12:34:56:78:9a:bc"; this applies only to clients, not servers. There was no option to disable this protection, so for the dataset to reflect an even more realistic traffic.
- The 46 features were split into two categories, namely, *Constant* and *Discrete*. That is, if a feature had constant values, it was scaled with the Min-Max technique. On the other hand, if the feature had discrete values, the One-Hot-Encoding (OHE) technique was used. Most of the features converted with the Min-Max scheme presented values with the scientific notation; these features were rounded to three decimal points. Some features, like the *quic.length* one, carried multiple values in the same frame. The respective values of these features were added into a single value.
- Due to the existence of multiple diverse protocols, namely TCP, UDP, QUIC, HTTP, and others, picking features from different protocols could result in having multiple empty cells. For this reason, we tinkered with the *Constant* features by replacing empty cells with 0, and with the *Discrete* ones by replacing empty cells with -1. This was done to clearly differentiate between the two states, i.e., *Constant*/empty vs. *Discrete*/empty.
- Finally yet importantly, we divided the 10 attacks into five classes, namely, *Normal, DDoS-flooding, DDoS-loris, Transport-layer*, and *HTTP/2* attacks, having the homonymous labels. In other words, the four attack classes reflect common characteristics of the respective included attacks, namely volume-based, "loris", "fuzzing", and HTTP/2-specific. Precisely, the *DDoS flooding* class comprises the HTTP3-flood, HTTP3-stream, and the QUIC-flood attacks. The *DDoS-loris* class consists of the HTTP3-loris and the QUIC-loris assaults. The *Transport-layer* class includes the Fuzzing and QUIC-enc, while the *HTTP/2 attacks* class contains the HTTP-smuggle, HTTP/2-concurrent, and HTTP/2-pause assaults.

For easy reference, the finally selected 46 features along with the utilized data preprocessing method per feature are given in Table 4. The left column designates the feature name as it was exported from *tshark*.

### 5.2. Experiments

For this initial evaluation of the dataset, we relied on commonly accepted ML techniques, without resorting to any optimization or dimensionality reduction schemes. Given that the dataset is imbalanced, the focus was on the AUC and F1 scores. Bear in mind that in the experiments, the 46-feature set of Table 4 was used.

#### 5.2.1. Shallow classification analysis

A number of common classifiers in the IDS domain were considered. For determining the optimal parameters, the *GridSearchCV* algorithm was used. *GridSearchCV* divides a dataset into x-fold partitions and evaluates the placed parameters for finding out the optimal ones. A 2-fold validation scheme was employed for evaluating the F1 score. The best results were obtained with the LightGBM, Decision Trees (DT), and Bagging classifiers. We did not use cross validation; the dataset was analyzed in a 60/40% fashion for the training and test sets, respectively. For equally splitting the dataset, the stratified split scheme was used.

**Table 4**
List of features used to evaluate the dataset.

| Feature name | Preprocessing |
| --- | --- |
| frame.len, ip.len, tcp.len, tcp.hdr_len, tcp.window_size_value, tcp.option_len, udp.length, tls.record.length, tls.reassembled.length, tls.handshake.length, tls.handshake.certificates_length, tls.handshake.certificate_length, tls.handshake.session_id_length, tls.handshake.cipher_suites_length, tls.handshake.extensions_length, tls.handshake.client_cert_vrfy.sig_len, quic.packet_length, quic.packet_number_length, quic.length, quic.nci.connection_id.length, quic.crypto.length, quic.stream.len, quic.token_length, quic.padding_length, http2.length, http2.header.length, http2.header.name.length, http2.header.value.length, http2.headers.content_length, http3.frame_length, http3.settings.qpack.max_table_capacity, http3.settings.max_field_section_size, dns.count.queries, dns.count.answers, http.content_length | Min-Max |
| tcp.flags.ack, tcp.flags.push, tcp.flags.reset, tcp.flags.syn, tcp.flags.fin, quic.long_packet_type, quic.fixed_bit, quic.spin_bit, quic.stream.fin, dns.flags.response, http.content_type | OHE |
| Label | – |

**Table 5**
Parameter values per Shallow classifier. A hyphen denotes that the current value is impertinent to the current ML algorithm.

| Parameters | DT | LightGBM | Bagging |
| --- | --- | --- | --- |
| max_depth | 200 | 1000 | – |
| max_leaf_nodes | 1000 | – | – |
| min_samples_leaf | 2 | – | – |
| min_samples_split | 10 | – | – |
| ccp_alpha | 0.0001 | – | – |
| max_bin | – | 2000 | – |
| min_child_samples | – | 30 | – |
| min_data_in_bin | – | 50 | – |
| min_split_gain | – | 0.1 | – |
| n_estimators | – | 1000 | 500 |
| num_leaves | – | 3000 | – |
| learning_rate | – | 0.01 | – |
| reg_alpha | – | 0.001 | – |
| reg_lambda | – | 0.001 | – |
| n_jobs | – | 1 | – |
| max_samples | – | – | 100,000 |

**Table 6**
Results on shallow classifiers.

| Model name | AUC | Prec. | Recall | F1 | Acc | T.t. |
| --- | --- | --- | --- | --- | --- | --- |
| DT | 76.67 | 65.93 | 63.56 | 64.21 | 94.44 | 00:04:30 |
| LightGBM | 77.49 | 79.71 | 64.72 | 68.40 | 94.76 | 05:42:05 |
| Bagging | 77.60 | 79.95 | 64.81 | 68.77 | 94.82 | 03:07:43 |

Table 6 presents the results per utilized classifier in terms of the AUC, Precision, Recall, F1-Score, and Accuracy (Acc) scores. The total time of each model's execution in hours/min/sec is also included in the rightmost column of the table. The Acc column is included just for the sake of completeness. As observed from the table, the best performer was the Bagging model with an AUC and F1 score of 77.60% and 68.77%, respectively. LightGBM performance was also very close to the Bagging one.

Fig. 7 depicts the confusion matrix of the top performer. The classifier missed ≈37K packets of the *Normal* class, misclassifying them as *DDoS-flooding*. Similarly, ≈105K packets of the *DDoS-flooding* class were wrongly identified as *Normal*. This indicates that the best performer experienced difficulties in differentiating between these two classes. An equivalent situation was perceived for the three remaining classes, namely, *Transport-layer, DDoS-loris*, and *HTTP/2 attacks*, where the algorithm missed ≈2.5K, ≈35K, and ≈7K packets, respectively, misplacing them to the *Normal* class. Overall, these results indicate that the best performer missed the samples of every attack-class in a percentage ranging between ≈25 to 75%, misplacing the corresponding samples to the *Normal* class.

### 5.2.2. DNN classification analysis

Three different models, namely, Multi-layer Perceptron (MLP), Denoising stacked Autoencoders (AE), and TextCNN were employed for DNN analysis. The parameters used per DNN model are recapped in Table 7. To obtain full control over the training phase, the mini-batch Stochastic Gradient Descent (SGD) optimizer was implemented, with a learning rate of 0.01 and a momentum of 0.9. Having said that, a low *Batch* size, e.g., 150 can result in a more generalized DNN model; this is because more data will be analyzed during each *Epoch*. In this respect, a *Batch* size of 256 was used. Moreover, where applicable, the well-known *ReLU* activator was utilized. Another common activator function for the output layer of DNN is the so-called *Softmax*. The latter was implemented to classify the results. No less important, a regularization effect was added through the *Dropout* scheme.

The input layer for TextCNN was diverse, namely the number of the dataset rows, while the output designated the number of classes, that is, five. Moreover, the same DNN model employed the Embedding layer of *Keras*, and has been utilized with three *Conv1D* hidden layers using the same padding. The *AveragePooling1D* layer was implemented after the first two hidden layers, while the *GlobalAveragePooling1D* was added after the third hidden layer. For both models, the *BatchNormalization* layer was applied after each hidden layer.

Additional techniques, including *Model Checkpoint* and *Early Stopping*, were applied to preserve the optimal training state of each DNN model. For these two schemes, we checked for the minimum loss value, and if the DNN model did not improve their loss value for two consecutive epochs, the training phase was ceased and the model was re-trained with the last optimal epoch. This eventually means that every fold was trained for a minimum two more epochs. These options, alongside other techniques, including *Dropout* and *validation test*, conceivably retained overfitting to the bare minimum.

The results in terms of the AUC metric per examined model are presented in Table 8. The penultimate column of the table indicates the number of epochs needed by the relevant DNN model to be trained. As observed from the table, in terms of the AUC metric, the AE and MLP models presented the best and worst performance scores, respectively (about 68.3% vs. 58.7%). Overall, the current scores lag behind vis-à-vis the scores yielded by shallow classification; nearly -9.30% and -15% for the AUC and F1 metrics, respectively. On the other hand, the TextCNN model was by far the fastest one for both set of features, requiring ≈5 hours. To provide a clearer picture of the results, Fig. 8 depicts the accuracy and validation performance of loss per epoch.

The above-mentioned inferior outcome is corroborated by the numbers in Fig. 9, which presents the confusion matrix of the top performer. It is easily perceived that, similar to the results of shallow classification, MLP experienced the same or even worse issues regarding the classification of the samples belonging to all the attack classes. Precisely, a higher percentage of samples of the *DDoS-*

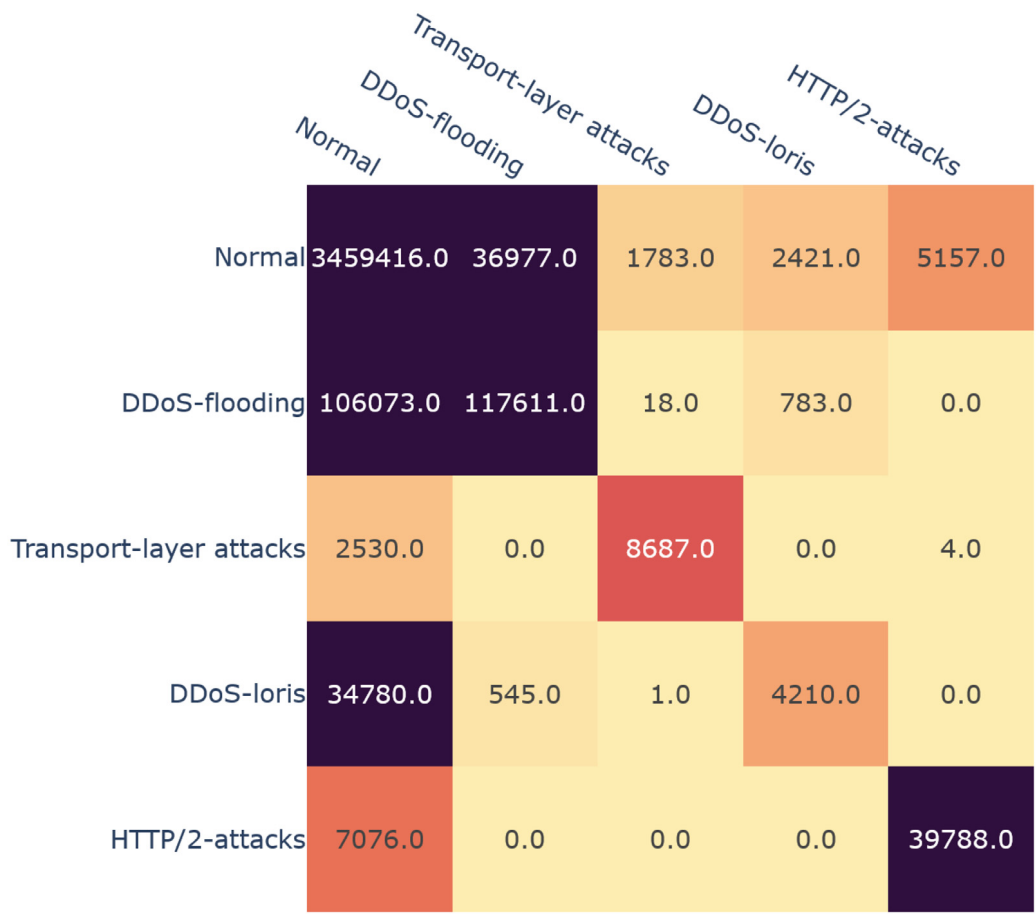

**Fig. 7.** Confusion matrix for the Bagging algorithm.

*flooding*, *DDoS-loris*, and *HTTP/2-attacks* classes – around 50%, 89%, and 15%, respectively – have been misplaced as *Normal* traffic.

### 5.2.3. Anomaly detection

Finally yet importantly, we analyzed the AE model through the anomaly detection method. While this model presented the worst results in Section 5.2.2, we chose it because, according to the literature, it is the commonest approach to anomaly detection. To this end, the dataset samples were divided into two classes, namely, *Normal* and *Malicious*. In a next step, using the stratified split scheme, the dataset was split into three subsets, i.e., *Train*, *Test*, and *Validation*, each having the 50%, 30%, and 20% of the dataset samples, respectively. The *Train* subset contained samples from only the *Normal* class, while both the *Test* and *Validation* subsets comprised samples from both classes. Therefore, the training phase was performed solely over samples of the *Normal* class.

The *Label* feature was removed from the *Test* subset, and kept to a separate subset, i.e., a *Label Test* subset. This was done to validate the results with the reconstruction error. The AE model was configured as in Table 7; the only difference was that the output layer contained one node because it had the *Linear* function as the output of the AE model. The loss function (SCC) was also replaced with the *Mean Absolute Error* (MAE). To make sure that the training phase did not suffer from overfitting, we compared the training loss against the validation loss values after each epoch. Indeed, as shown in Fig. 10, the training phase did not exhibit overfitting. The model was trained for 13 epochs, and produced a MAE of 0.3228 after the last epoch.

**Table 7**
Parameter values per DNN algorithm. A value of "/3" or "/4" in the MLP Dropout parameter indicates the number of layers in which this parameter received the designated value. The layer values are calculated without including the input and output ones. SCC stands for Sparse Categorical Crossentropy. A hyphen denotes a non-applicable option for this DNN model.

| Parameters | MLP | Autoencoders | TextCNN |
| --- | --- | --- | --- |
| Activator | ReLU | ReLU | ReLU |
| Output activator | Softmax | Softmax | Softmax |
| Initializer | He_uniform | – | – |
| Optimizer | SGD | SGD | SGD |
| Momentum | 0.9 | 0.9 | 0.9 |
| Dropout | 0.25/3-0.2/4 | 0.25 | 0.2 |
| Learning rate | 0.01 | 0.01 | 0.01 |
| Loss | SCC | SCC | SCC |
| Batch Norm. | ✓ | ✓ | ✓ |
| Embedding layer | ✗ | ✗ | ✓ |
| Hidden layers | 7 | 12 | 4 |
| Nodes (Per layer) | 300/200/160/120/60/30/10 | 300/200/160/120/60/30/10/30/60/120/160/200/300 | Conv1D(64,12)/Conv1D(128,10)/Conv1D(256,12) Dense(200) |
| Batch size | 256 | 256 | 256 |

**Table 8**
Results of DNN model analysis.

| Model name | AUC | Prec. | Recall | F1 | Acc | Epochs | T.t. |
|---|---|---|---|---|---|---|---|
| MLP | 68.34 | 62.15 | 51.15 | 53.71 | 92.62 | 68 | 07:16:34 |
| AE | 58.72 | 63.50 | 33.15 | 39.59 | 92.51 | 48 | 08:19:29 |
| TextCNN | 64.52 | 72.44 | 45.9 | 46.71 | 92.37 | 42 | 05:12:10 |

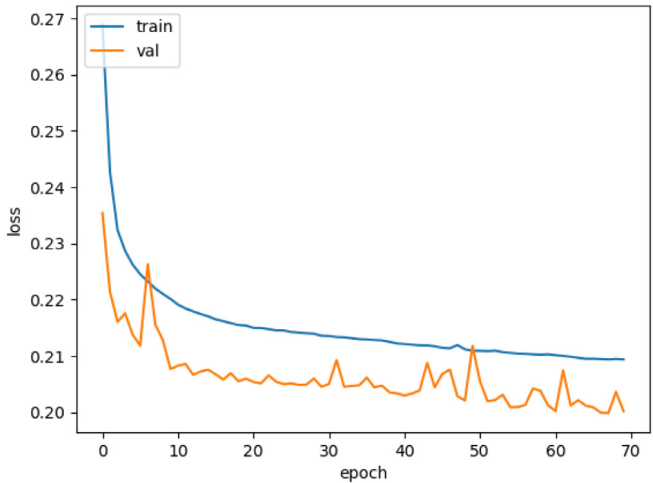

**Fig. 8.** MLP model comparison of training loss with the validation loss, after each epoch.

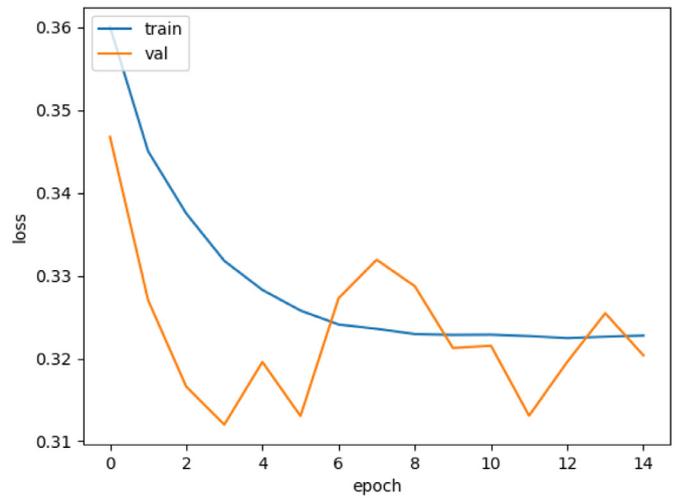

**Fig. 10.** Anomaly detection model comparison of training loss with the validation loss, after each epoch.

After the training phase, the model was evaluated by calculating the reconstruction error. For this purpose, first, the model was requested to predict the *Test* subset. Then, the MAE was computed by taking the absolute difference between the derived predictions and the *Test* subset values. To identify which of these predictions was correctly chosen, i.e., to compare and separate the MAE error, the *Label* subset was used. Precisely, if the label for a sample was *Malicious*, the corresponding MAE error was flagged as an anomaly, while the remaining samples were flagged as *Normal*. By trial and error, it was calculated that the threshold for this analysis should be 0.33. Fig. 11 depicts this observation, highlighting that the model was mostly unable to discern between the two classes. This means that a better feature selection process is probably needed, as already mentioned in Section 5.2.2.

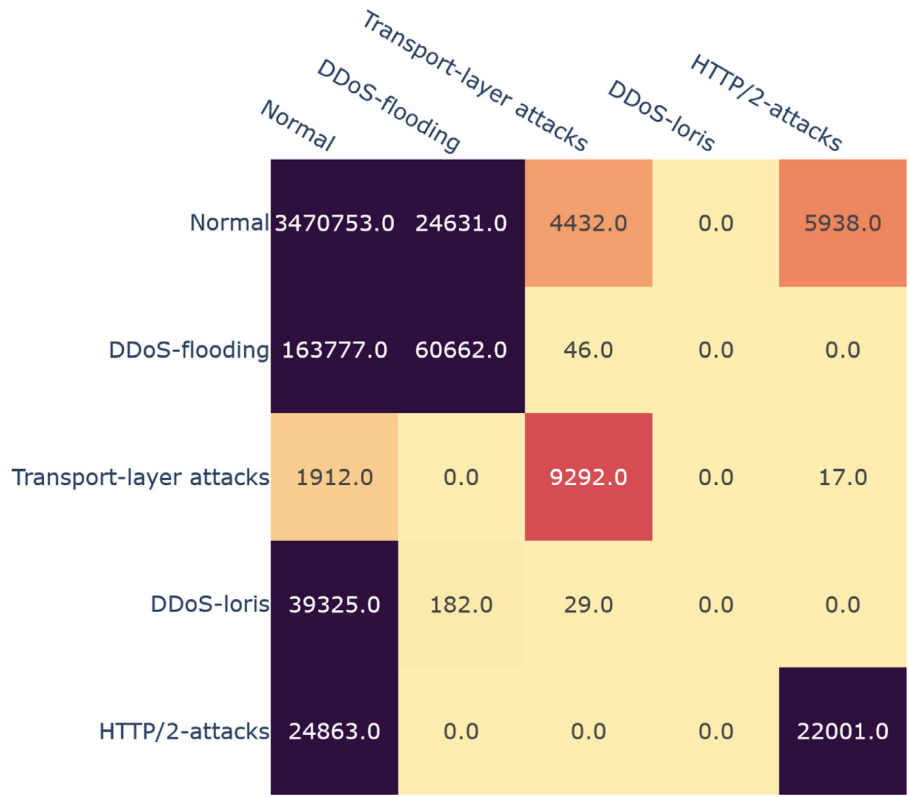

**Fig. 9.** MLP confusion matrix.

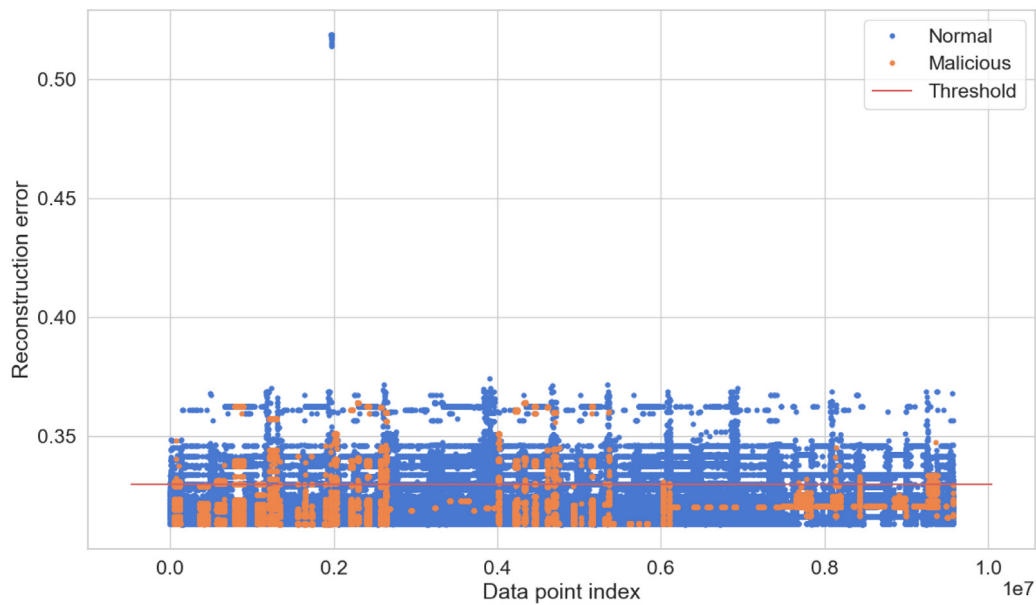

**Fig. 11.** Anomaly detection model – reconstruction error with threshold.

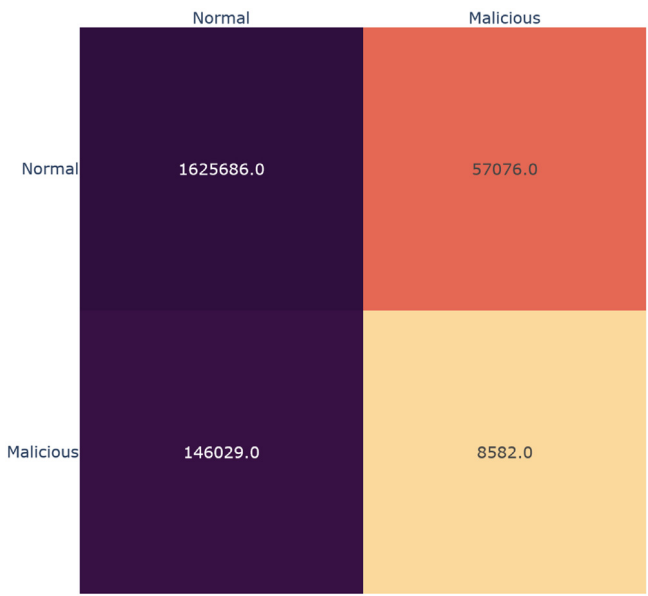

**Fig. 12.** Anomaly detection model confusion matrix based on threshold.

Moreover, Fig. 12 illustrates the prediction rate of the current model. For generating this confusion matrix, we labeled every sample above the threshold as *Malicious*, while the rest of the samples were marked as *Normal*. As seen from the figure, the results are imprecise, i.e., the *Malicious* class is largely confused with the *Normal*. On the other hand, the latter class misplaced a rather small percentage of the samples (≈3.4%) as *Malicious*. Regarding legacy evaluation metrics, namely, Precision, Recall, F1-score, and Acc, they presented tolerable results, i.e., 91.75%, 96.60%, 94.11%, and 88.94%, respectively. Obviously, this behavior, i.e., a lower Acc score vis-à-vis F1, is due to the imbalanced nature of the dataset.

## 6. Conclusion

The work at hand delivers the first to our knowledge full-fledged study on HTTP/2 security, extending the identified attacks to its successor, namely HTTP/3. In Section 2 we reviewed the literature for HTTP/2 attacks, in Section 3 we considered attacks that have not already been evaluated (i.e., in the same testbed in Chatzoglou et al. (2022d)), and in Section 4 together with the attacks of the previous sections we included additional ones to create a blend of HTTP/2, HTTP/3, and QUIC assaults. In Table 9, we demonstrate the presence of each attack in each of the three sections of our paper dealing with attacks (i.e., Section 2:Taxonomy, Section 3:Hands-on evaluation, Section 4:Dataset). In the first column of the table we present all the attacks. The second column shows which of these attacks were considered in the taxonomy of HTTP/2 attacks or the reason why they could not be considered. The third column shows which attacks could be considered in the hands-on evaluation against HTTP/3 and which not; in the latter case a proper justification is given. Finally, in the fourth column we present the attacks considered in the dataset.

Regarding our methodology, starting with a review of HTTP/2 attack categories, we examine half a dozen of contemporary HTTP/3-enabled servers regarding their resilience against either common or uncommon attack tactics. This endeavor yielded interesting results, some of which leading to CVE. What is more, through the creation of a realistic testbed, we created a rich, voluminous (30 GB) dataset containing an assortment of 10 attacks against HTTP/2, HTTP/3, and QUIC. The dataset, coined "H23Q", is labeled and is offered publicly to the community. A preliminary evaluation of the dataset conducted by means of different techniques on a set of 46 cross-layer features revealed, as expected, that certain attack classes are very challenging to detect. In this respect, future work can concentrate on both cherry-picking of more informative features and the use of more sophisticated IDS tech-niques, including network flow analysis and time series anomaly detection.

**Table 9**
An overview of the attacks presented in our paper.

| Attack | Addressed in section | | |
|---|---|---|---|
| | Section 2 (Taxonomy) | Section 3 (Hands-on evaluation) | Section 4 (Dataset) |
| HTTP/3 flooding | ✓ | ✓ | ✓ |
| Slow-rate HTTP/3 | ✓ | ✓ | ✓ |
| HTTP/x downgrade | ✗: passive attack; the smuggling attack below has a similar effect | ✓ | ✗: passive attack; the smuggling attack below has a similar effect |
| HTTP/3-tables/streams | ✗: not an HTTP/2 attack | ✓ | ✓ |
| quic-flooding | ✗: not an HTTP/2 attack | ✗: hands-on evaluation performed in Chatzoglou et al. (2022d) | ✓ |
| quic-encapsulation | ✗: not an HTTP/2 attack | ✗: hands-on evaluation performed in Chatzoglou et al. (2022d) | ✓ |
| quic-loris | ✗: not an HTTP/2 attack | ✗: hands-on evaluation performed in Chatzoglou et al. (2022d) | ✓ |
| quic-fuzz | ✗: not an HTTP/2 attack | ✗: hands-on evaluation performed in Chatzoglou et al. (2022d) | ✓ |
| Smuggling | ✓ | ✗: it has a similar effect with the HTTP/x downgrade attack above | ✓ |

Other possible avenues for future work include: (i) the analysis of HTTP/2 Websockets McManus (2018) from a security perspective; note that the bootstrapping of WebSockets with HTTP/3 is just around the corner Hamilton (2022), and (ii) the development of a full-featured HTTP/x fuzzer, enabling meticulous vulnerability testing. Thus far, the only HTTP/2-focused fuzz tool is the so-called *http2fuzz* Larsen and Villamil (2015), which however is quite outdated.

**Declaration of Competing Interest**

The authors declare that they have no known competing financial interests or personal relationships that could have appeared to influence the work reported in this paper.

**CRediT authorship contribution statement**

**Efstratios Chatzoglou:** Conceptualization, Methodology, Investigation, Writing – original draft, Data curation, Validation, Software. **Vasileios Kouliaridis:** Investigation, Writing – original draft, Writing – review & editing. **Georgios Kambourakis:** Conceptualization, Methodology, Validation, Writing – original draft, Writing – review & editing, Supervision. **Georgios Karopoulos:** Methodology, Writing – original draft, Writing – review & editing. **Stefanos Gritzalis:** Supervision, Writing – review & editing.

**Data availability**

The HTTP/3 attacks are available in the following repository: https://github.com/efchatz/HTTP3-attacks. The "H23Q" dataset is available for download at the well-known AWID website at https://icsdweb.aegean.gr/awid/other-datasets/H23Q.

**Efstratios Chatzoglou** received the MSc degree in security of information and communication systems from the University of Aegean, Samos, Greece. He previously worked as a Web Developer and a Penetration Tester at the Hellenic National Defence General Stuff. He is currently a Penetration Tester at TwelveSec and a research associate at Info-Sec-Lab with the University of Aegean. His research interests lie in the fields of wireless and cellular networks security, IoT networks security, Android application security, Web application security, and Machine Learning.

**Vasileios Kouliaridis** received his PhD from the dept. of Information and Communication Systems Engineering, University of the Aegean, Greece. His research interests are in the areas of Android security and privacy, mobile malware analysis and detection, and machine learning. He has published and is a frequent reviewer in conferences and scientific journals in the above areas.

**Georgios Kambourakis** is a full professor at the dept. of Information and Communication Systems Engineering, University of the Aegean, Greece. He has served as the head of the dept. from Sept. 2019 to Oct. 2019, and was the director of Info-Sec-Lab from Sept. 2014 to Dec. 2018. His research interests are in the fields of mobile and wireless networks security and privacy, VoIP security, IoT security and privacy, DNS security, and security education, and he has more than 160 refereed publications in the aforementioned areas. More info at: http://www.icsd.aegean.gr/gkamb.

**Georgios Karopoulos** is a Scientific Officer at the Joint Research Centre of the European Commission. He holds a Ph.D. degree in computer network security from the University of the Aegean, Greece. In the past, he was a Marie Curie fellow researcher at the University of Athens, Greece, and an ERCIM fellow at IIT-CNR, Italy. His research interests are in the areas of network security, smart grid security and critical infrastructure protection. He has published and he is a frequent reviewer in conferences and scientific journals in the above areas.

**Stefanos Gritzalis** is a Professor of Information and Communication Systems Security, at the Lab. of Systems Security, Dept. of Digital Systems, University of Piraeus, Greece (2019+) and Director of the Postgraduate Programme "MSc in Law and Information & Communication Technologies" (2020+). He is a Member of the Board of the Hellenic Authority for Communication Security and Privacy (2020+). He was the Rector of the University of the Aegean, Greece (2014-2018). He has acted as Special Secretary for the Hellenic Ministry for Administrative Reform and Electronic Governance. Previously, he was a Professor at the University of the Aegean, Greece, School of Engineering, Dept. of Information and Communication Systems Engineering, and member of the Info-Sec-Lab Laboratory of Information and Communication Systems Security. He was the Head of the Dept. of Information and Communication Systems Engineering, Deputy Head of the Dept. of Information and Communication Systems Engineering, and Director of the Lab. of Information and Communication Systems Security. He holds a BSc in Physics, an MSc in Electronic Automation, and a PhD in Information and Communications Security from the Department of Informatics and Telecommunications, University of Athens, Greece. His published scientific work includes more than 11 books and 37 book chapters. His work has been published in 325 papers (142 in refereed journals and 183 in the proceedings of international refereed conferences and workshops). The focus of his publications is on Information and Communications Security and Privacy. He acts as Area Editor for the prestigious "IEEE Communications Surveys and Tutorials" journal. He is the Editor-in-Chief or Editor or Editorial Board member in 35 journals and a Reviewer in more than 80 scientific journals. More info at: https://www.ds.unipi.gr/en/faculty/sgritzalen/.